\documentstyle[epsf]{elsart}

\def\sla#1{\rlap\slash #1} 

\begin{document}
\vspace*{-1cm}
\normalsize
\hspace*{6.5cm}\parbox{15cm}{Accepted in Nucl.Phys.{\bf A} (1999)}
\begin{frontmatter}
\title{Covariance of Light-Front Models: Pair Current} 
\author[PARIS]{J.P.B.C. de Melo}, 
\author[ITA]{T. Frederico}, 
\author[ITP,TNO]{H.W.L. Naus} 
\author[ITP]{P.U. Sauer}
\address[PARIS]{Division de Physique Th\'eorique, 
Institut de Physique Nucl\'eaire,
F-91406 Orsay Cedex, France  and Laboratoire de Physique Th\'eorique et
Particules El\'ementaires, Universit\'e Pi\'erre \& Marie Curie,  
4 Place Jussieu, F-75252 Paris Cedex 05, France } 
\address[ITA]{Departamento de F\'\i sica, ITA, Centro T\'ecnico
Aeroespacial, 12.228-900, S\~ao Jos\'e dos Campos, Brazil}
\address[ITP]{Institute for Theoretical Physics, University Hannover \\ 
D-30167 Hannover, Germany}
\address[TNO]{TNO Physics and Electronics Laboratory,\\ P.O. Box 96864,
2509 JG The Hague,
The Netherlands}
\date{\today}

\begin{abstract}
We compute the + component, {\it i.e.},
$j^+=j^0+j^3$, of the electromagnetic current of a composite
spin-one two-fermion system
for vanishing momentum transfer component $q^+=q^0+q^3$.
In particular, we extract the nonvanishing
pair production amplitude on the light-front. 
It is a consequence of the longitudinal zero momentum mode, 
contributing to the light-front  current in the Breit-frame. 
The covariance of the current is violated, if such pair terms 
are not included in its matrix elements. We illustrate our discussion
with some numerical examples.
\end{abstract}

\vspace{ 1cm}

\hspace{ 2cm}{\bf PACS:} 12.39.Ki,14.40.Cs,13.40.Gp

\end{frontmatter}
\vskip 0.2cm

% \twocolumn
% \narrowtext

%%  \newpage

\section{ Introduction}

The light-front hypersurface
given by $x^+=t+z=0$ (null-plane) is adequate for
defining a light-front wave-function, as suggested by Dirac \cite{dirac}.
The null-plane is invariant under seven kinematical transformations,
three of them include boosts. It is possible to keep a limited 
covariance of physical observables under kinematical boosts, 
within a fixed number of degrees of freedom \cite{wein}. 
However, the rotations around the $x$- and $y$-axis do not belong to the 
stability group; the corresponding transformations are
not kinematic and therefore particle-antiparticle pair creation is possible. 
Thus, the description of a physical system in terms of a
composite null-plane wave-function with a minimal, fixed number of particles
in general violates complete covariance.

The light-front electromagnetic (e.m.) current $j$ of a
composite system can in principle be  constructed from  covariant Feynman 
diagrams. In the Breit-frame with  momentum transfer $q^+=q^0+q^3=0$,
the one-loop expressions can be 
integrated over the  $k^- \ (k^-=k^0-k^3)$ component
of the internal loop momentum \cite{saw92,tob92,pach97,pion99}.
This procedure can be compared to the exact relation of the "old-fashioned" 
perturbation theory with the covariant perturbative expansion
of the field theoretical S-matrix. In a series of papers by
Chang, Root, and Yan \cite{yan}, it was demonstrated that the
perturbative expansion of light-front field theory is obtained from the
covariant Feynman amplitudes by first integrating over  $k^-$
in the momentum loops. However, in some cases the naive 
Cauchy integration of the $k^-$ momentum 
gave completely wrong results \cite{yan}.
The connection between light-front and covariant field theory
was  further developed in the recent works 
by Ligterink and Bakker \cite{ba95} and
Schoonderwoerd and Bakker \cite{ba98}, with special attention
to the separation of the zero-mode in the perturbative amplitudes.
The failure of the naive Cauchy integration amounts to
neglecting the longitudinal 
zero-momentum modes \cite{pion99,ba95,pach98,rik98,ji98}. 

In this paper, we discuss the current of a vector particle, treated as
a composite fermion system. We explicitly calculate the Feynman one-loop
triangle diagram for two different couplings of the fermions to the vector
particle. In \cite{pach97} 
it has been shown  that
the current component $j^+$ of this  spin-one  system violates
the requirements of covariance, parity and gauge invariance, 
the so called "angular condition" \cite{inna84,to93,keis94,inna95},
 when the naive Cauchy integration
is performed in the triangle diagram for $q^+=0$. 
A complete elimination of the
pairs created out or annihilating into the vacuum was believed to be possible,
leading to a description in terms 
of a two-particle light-front wave-function 
\cite{saw92,tob92,pach97,pion99}.
Although such a current is covariant
under kinematical boost transformations, it does not have the
correct transformation properties 
under general rotations and parity transformations. It has been 
pointed out in an example with $\gamma^\mu$-coupling, that this problem 
appears once a nonvanishing pair term around $q^+=0$ 
in the $k^-$ integration is neglected \cite{pach98}. 

Therefore we focus in this work on the difficulty in the
$k^-$ integration and its cure by a careful treatment of singularities.
Our treatment is different and the result is more general than the one in 
Refs. \cite{ba95}.
In the  case of the $j^+$ component,
the longitudinal zero-mode contribution  to the amplitude 
is equal to the one 
for  virtual pair creation by the photon with  $q^+=0$  \cite{pach98}.
These light-front pair production 
contributions to the $j^+$ current of a vector composite 
particle at $q^+=0$ are derived using the
method of "dislocation of pole integration", developed 
in Refs. \cite{pion99,pach98,rik98}.  
The contributions originate from
the instantaneous part of the light-front fermion propagator
and from the derivative coupling of the composite
particle to the fermions. We use the covariant model 
proposed in Ref.\cite{pach97} for the rho-meson  e.m.
current, in which Pauli-Villars regulators are introduced 
in the triangle diagram. For comparision we also perform
the covariant calculation which corresponds to directly integrating the 
momentum loop in the standard variables of the
four-dimensional phase-space, {\em i.e.}, first integrating over
  $k^0$ analytically and
then numerically in the remaining three-dimensional momentum space.

This paper is organized as follows. In  the next section
the electromagnetic current of a spin-one  particle  
is discussed. In section III the method of dislocating poles 
is outlined. The model for the electromagnetic current is  presented in 
section IV and the importance of including  the pair terms 
in the light-front calculation is emphasized. 
In the last section  
we show our numerical results, discuss their implications
 and conclude with a brief summary.

\section{Electromagnetic current}

The general expression of the e.m. current of a
 spin-one particle has the form \cite{frank} :
 \begin{eqnarray}
j_{\alpha \beta}^{\mu}=\left[ F_1(q^2)g_{\alpha \beta}-F_2(q^2)
\frac{q_{\alpha}q_{\beta}}{2 m_v^2}\right] (P'+P)^\mu-
F_3(q^2)\left[q_\alpha g_\beta^\mu-
q_\beta g_\alpha^\mu\right] \ ,
\label{eq:curr1}
\end{eqnarray}
where $m_v$ is the  mass of the composite  vector particle, $q^\mu$ the
momentum transfer and 
 $P^\mu$(${P'}^{\mu}$) is  the initial (final) on-mass shell momentum.

The impulse approximation of the electromagnetic current 
$J^{\mu}$ is given by \cite{pach97}:
\begin{eqnarray}
 j^\mu_{\alpha \beta}&=&\imath  \int\frac{d^4k}{(2\pi)^4}
\Lambda(k,P')\Lambda(k,P)
\nonumber \\
& &\times
\frac{ Tr[\Gamma_\alpha '
(\sla{k}-\sla{P'} +m) \gamma^{\mu} 
(\sla{k}-\sla{P}+m) \Gamma_\beta  
(\sla{k}+m)]}
{((k-P)^2 - m^2+\imath\epsilon) 
(k^2 - m^2+\imath \epsilon)
((k-P')^2 - m^2+\imath \epsilon)}  
\ , 
\label{jvec}
\end{eqnarray}
where $\Gamma '_\alpha$ and $\Gamma_\beta$ define the coupling of the
fermions to the vector particle.
The functions $\Lambda $ are used to regularize the model, {\it i.e.},
render the momentum integrals finite. These Pauli-Villars
regulators, was well as the couplings, 
will be explicitly defined in section IV.

The matrix elements  
\begin{eqnarray}
j^+_{ji}:= \epsilon_j^{'\alpha}j^+_{\alpha\beta}\epsilon_i^{\beta}
\label{mjp}
\ ,
\end{eqnarray}
in the instant-form spin basis are obtained from Eq.(\ref{eq:curr1}) 
in the Breit-frame, where $q^\mu=(0,q_x,0,0)$. The initial
 instant-form cartesian polarization four-vectors 
of the spin-one particle are
\begin{eqnarray}
\epsilon^\mu_x=(-\sqrt{\eta},\sqrt{1+\eta},0,0) \ , \
\epsilon^\mu_y =(0,0,1,0) \ , \
\epsilon^\mu_z = (0,0,0,1) \ , \
\label{eq:pol} 
\end{eqnarray}

with $\eta=  - q^2/4 m_v^2 $.
For the final polarization states we have
\begin{eqnarray}
\epsilon_x^{'\mu}= (\sqrt{\eta},\sqrt{1+\eta},0,0) \ , \
\epsilon_y^{'\mu} = \epsilon^\mu_{y} \ , \
\epsilon_z^{'\mu}= \epsilon^\mu_{z} \ . \
\label{eq:polp}
\end{eqnarray}

The vector particle four-momenta   are 
$P^\mu=(P^0,-q_x/2,0,0)$ for the initial state 
and  ${P'}^\mu=(P^0,q_x/2,0,0)$ for the final state;
  $P^0=m_v \sqrt{1+\eta} $. In this frame, the nonzero
matrix elements of  $j^+$ 
 are $j^+_{xx}$,  
$j^+_{zx}\ (=-j^+_{xz})$, $ j^+_{yy}$ and $ j^+_{zz}$.
The angular condition corresponds to the equality 
$ j^+_{yy}\ = \ j^+_{zz}$ \cite{to93} and should be satisfied for
the current matrix elements, Eq.(\ref{mjp}), 
for the particular polarization vectors given above.

In the next section, we will present the analytical 
method that allows to integrate the particular 
terms of the impulse approximation of the current, Eq.(\ref{jvec}), 
which contain the amplitude for pair production by the incoming photon.

\section{ The "dislocation of pole" integration}

Here we explicitly consider the
one-loop integration over  the $k^-$ momentum  variable
for arbitrary number of propagators and 
Pauli-Villars regulators \cite{pach98,rik98}. 
We generalize the method that  was  derived to compute the pair production
amplitude for the $j^-$ component of the current of a composite boson
in the limit of $q^+\rightarrow 0$ \cite{pach98}.
The technique has also been succesfully
applied in the  context of a perturbative bosonic model in light-front
field theory \cite{rik98}. In a recent work, we have used this method 
in the calculation of both components, $j^{+}$ and $j^{-}$, of the
electromagnetic current of the pseudoscalar pion.  The relevance of pair terms
for the $j^{-}$ component of this  current was demonstrated \cite{pion99}.

The integral in the variables $k^+$ and $k^-$
is given by the building blocks
\begin{eqnarray}
I^{(N)}_{mn}(\{ f_i\} )=\int
\frac{d k^{+} d k^- }{2} 
\frac{(k^-)^m(P^+-k^+)^n}{\prod_{j=1}^N(P^+ -k^+)
(P^- - k^-- \frac{f_j -\imath \epsilon}{P^{+}-k^+})} \ ,
\label{imn}
\end{eqnarray}
with external four-momentum components $P^+$ and $P^-$ and $f_j >0$.
The parameters $m, \ n$  are limited  to be
 $m\leq n$; the other integer $N$ should be large enough to
render the integral finite.
Note that Eq.(\ref{imn}) would be zero by naive Cauchy integration over $k^-$.
As we will prove, $I^{(N)}_{mn}(\{ f_i \})$ does not depend on 
the specific values of $P^+\ > \ 0$ and $P^-$.

In the denominator  one of the $N$ poles
is dislocated via
$P^+\rightarrow P^+ + \, \delta  \;\;(\delta > 0) $, {\it i.e.},
\begin{equation}
\frac{1}{(P^+-k^+)(P^--k^-)  - f_1 +i \epsilon} \rightarrow
\frac{1}{(P^+ + \delta-k^+)} \,
\frac{1}{ P^- -k^-- \frac{f_1 -i \epsilon}{P^+ \, + \, \delta -k^+}} \,  ,
\label{dis}
\end{equation}
violating the symmetry with respect to the permutation
of the factors $j$. However, it is shown below that the final  results
are symmetrical under permutation of $f_j$.

Nonanalytical terms in $\delta$ are not present after the 
 $k^-$ integration. Thus, the limit  $\delta \rightarrow 0$ 
can be performed. Either the integration vanishes with 
positive integer power of $\delta$  or
the infinitesimal parameter $\delta$ can be eventually  absorbed by 
rescaling  the variable $k^+$ 
($x=(k^{+}-P^{+})/\delta)$, as done 
in \cite{pion99,pach98,rik98}. 
The limit $\epsilon \rightarrow 0$ can also  be taken and
the results are:
\begin{eqnarray}
I^{(N)}_{mn}(\{ f_i \} )=0 \ ,  \ \ \ m \ <\ n \ ,
\label{i1}
\end{eqnarray}
and
\begin{eqnarray}
I^{(N)}_{mm}(\{ f_i\} )=\imath \pi \sum_{j=1}^N\frac{ (-f_{j})^m\ln(f_{j})}
{\prod_{k=1,k\neq j}^N(f_j - f_k)} \ , \ \ \ m=n \ .
\label{imm}
\end{eqnarray}
The first result, Eq.(\ref{i1}), immediately follows from Cauchy's theorem.
The results contained in  Eq.(\ref{imm})  are proved in the Appendix.
The  case of $I^{(N)}_{00}(\{ f_i\} )$ has also
been obtained in Ref.\cite{ba95} using a different technique. 
The form of Eq.(\ref{imm}) is again symmetrical under permutation of $f_j$,
a posteriori justifying our choice in Eq.(\ref{dis}).

Using a model for the composite vector particle \cite{pach97},
we will show that a naive, incorrect evaluation of
the integrals $I^{(N)}_{mn}(\{ f_i \} )$ causes the seeming violation of
 covariance in the form-factor calculation using
$j^+$ in a frame with vanishing momentum transfer component, $q^+=0$. 
It is necessary to include pair production, implicitly
given by Eq.(\ref{imm}), in order to get agreement 
with the covariant calculation. The interpretation of  results obtained with
the ``dislocation of  pole'' method, especially the identification of
zero mode contributions as pair terms, was given
in Refs.\cite{pion99,pach98,rik98}.

\section{The model and discussion of pair terms}

In order to  discuss the
"good" component $j^+$ of the current \cite{dash} of the
vector particle, we specify the covariant model 
\cite{pach97}. The composite spin-one particle is described as
a two-fermion system and two forms of the coupling in Eq.(\ref{jvec})
have been used:
\begin{eqnarray}
i)\ \Gamma_\mu =\Gamma_\mu'= \gamma_\mu \ , \ \ ii)\ 
\Gamma_\mu=(2k_\mu-P_\mu) \ ;
\Gamma_\mu'=(2k_\mu-P'_\mu) \ ,\label{eq:ver}
\end{eqnarray}
where the primed quantities refer to the final state.
For the regularization function $\Lambda(k, P)$, choosen to make
the integral of Eq.(\ref{jvec}) finite, we take
$
\Lambda(k,P) =(2\pi)^2C\left[(k-P)^2 - m^2_R+\imath\epsilon\right]^{-2}
$.  
This choice serves our purpose
of presenting the main points on how the pair current preserves the
rotational invariance of $j^+$  on the light front.
It also allows to
identify a null-plane wave-function, when one looks at contribution
of the residue due to the pole of the spectator particle 
for $k^-=\frac{k^2_\perp+m^2-i \epsilon}{k^+}$ with $0 < k^+ <P^+$ in  
the $k^-$ integration of Eq.(\ref{jvec}) \cite{pach97}. The 
factor $C$ is fixed by the charge normalization.
The model fulfills current conservation
\cite{pach97}.

The instantaneous part of the fermion propagator
is the second term of the Dirac
propagator decomposed using the light-front momenta
\begin{eqnarray}
\frac{\sla{k}+m}{k^2-m^2+\imath \epsilon}=
\frac{\sla{k}_{(on)}+m}{k^+(k^--k^-_{(on)}+\frac{\imath \epsilon}{k^+})}
+\frac{\gamma^+}{2k^+} \ ,
\label{inst}
\end{eqnarray}
where $k^-_{(on)}=\frac{k_\perp^2+m^2}{k^+} $.
The presence of a nonvanishing zero-mode contribution to  $j^+$ 
is, in case of  $\gamma_\mu$-coupling  Eq.(\ref{eq:ver} $i$)
only due to the instantaneous part of the Dirac propagator. 
For  derivative coupling  Eq.(\ref{eq:ver} $ii$) the zero-mode contribution
is caused by the vertices as well.

Eq.(\ref{jvec}) yields the  matrix elements
of the current depending on the polarization states.
We call the matrix elements of $j^+$,  which
 do not couple to the zero-mode 
and  correspond to the numerator of Eq.(\ref{jvec}) being
dependent only on $k^+$, $k_\perp$ and 
$(k^-)^{m+1}(P^+-k^+)^n$ with $m < n$ in Eq.(\ref{i1}),
``good'' matrix elements. 
In such cases only the pole at  which the spectator
particle is on-mass-shell,  contributes to the Cauchy 
integration over $k^-$ in the limit of $\delta\rightarrow 0_+$.
The pole in Eq.(\ref{jvec})
is placed on the lower
half of the  complex $k^-$ plane at $ k^-=(k^2_\perp+m^2-\imath
\epsilon)/k^+$ with $0< k^+ <P^+$.
\vspace{0.2cm}\\
{\bf i) $ \gamma^\mu$-coupling} 

The instantaneous component of the fermion propagator, cf. 
Eq.(\ref{inst}),
yields a nontrivial zero-mode  coupling to $j^+$. This contribution can be
separated in Eq.(\ref{jvec}) and can be written in the following form 
\begin{eqnarray}
j^{+pair}_{xx}&=& -\eta B_\gamma(q^2) \ , \
\nonumber \\
j^{+pair}_{zx}&=&  - \sqrt{\eta} B_\gamma(q^2) \ , \
\nonumber \\
j^{+pair}_{zz}&=&   B_\gamma(q^2)
\ ,
\label{pairv}
\end{eqnarray}
where
\begin{eqnarray}
B_\gamma(q^2)=4\imath  C^2\int \frac{d^2k_\perp}{P^+} I^{(6)}_{00}(\{ f_i \} )
\left( k_\perp^2 + m^2   + \frac{q^2}{4} \right)
 \ , 
\label{bvec}
\end{eqnarray}
with the functions $f_i$ given by
\begin{eqnarray}
&& f_1  =  (P-k)^{2}_\perp+m^2 \ ;  \
f_2  =  (P'-k)^{2}_\perp+m^2 \ ; \ 
f_3 = (P-k)^{2}_\perp+m_{R}^2 \ ; \
\nonumber \\
&& f_4  = (P'-k)^{2}_\perp+m_{R}^2 \ ; \
f_5=f_3 \ ; \
f_6=f_4 \ .
\label{f}
\end{eqnarray}
\vspace{0.2cm}\\
{\bf ii) Derivative coupling} 

In this example, the virtual pair production amplitude by the
$q^+\rightarrow 0$ photon  has 
two sources. The first one is the instantaneous part of the
fermion propagator and the second one is the light-front time
derivative in the vertex.  The numerator in the expression for
$j^+$, cf. Eq.(\ref{jvec}) and Eq.(\ref{mjp}), is written as:
\begin{eqnarray}
&& Tr\left[\Gamma_\mu '{\epsilon{'}_j^{\mu}}
(\sla{k}-\sla{P'} +m) \gamma^{+} 
(\sla{k}-\sla{P}+m) \Gamma_\nu  {\epsilon_i^{\nu}}
(\sla{k}+m)\right]=
\nonumber \\
&&(2k-P')_\mu {\epsilon{'}_j^{\mu}}
(2k-P)_\nu  {\epsilon_i^{\nu}}
Tr\left[ (\sla{k}-\sla{P'} +m) \gamma^{+} 
(\sla{k}-\sla{P}+m) (\sla{k}+m) \right] . 
\label{trder}
\end{eqnarray}
The pair production amplitude can now be found from  Eq.(\ref{trder}). We first 
collect the terms that depend on powers of $k^-$, use
 $I^{(N)}_{mn}=0$ for $m \ < \ n $, and then insert the identity
\begin{eqnarray}
\frac{(k^-)^2k^+}{k^2-m^2+i\epsilon}=
(k^--P^-)+P^-+ \frac{ k_\perp^2 +m^2}{k^+}
+\frac{k^-_{(on)}( k_\perp^2 +m^2)}{k^2-m^2+i\epsilon}
\ .
\label{eq:id}
\end{eqnarray}
This identity is used, whenever the factor 
$(k^-)^2k^+$ appears without being multiplied by powers of $(P^+-k^+)$. The
first term in the r.h.s of Eq.(\ref{eq:id}) is odd under the transformation
$(P^--k^-)\rightarrow \ -(P^--k^-)$ and thus its contribution 
to the pair current is zero. The last term of the r.h.s. of Eq.(\ref{eq:id})
also does not contribute to the pair current. The final
result is
\begin{eqnarray}
j^{+pair}_{xx}&=&\eta B_D(q^2) \ ,
\nonumber \\
j^{+pair}_{zx}&=& \sqrt{\eta}B_D(q^2) \ ,
\nonumber \\
j^{+pair}_{zz}&=&- B_D(q^2) \ ,
\label{pairder}
\end{eqnarray}
where $B_D(q^2)$ is given in terms of the functions $f_i$, cf. Eq.(\ref{f}),
\begin{eqnarray}
B_D(q^2)&=&
4\imath C^2\int \frac{d^2k_\perp}{P^+}
\{
I^{(6)}_{22}(\{ f_i \} ) +2 I^{(6)}_{11}(\{ f_i\} ) (k_{\perp}^{2}+m^2) +
\nonumber \\
& &I_{00}^{(6)}(\{ f_i\} )\left( k_{\perp}^2+m^2+\frac{q^2}{4}\right)
\left( k_{\perp}^2+m^2-m_{v}^2 (1+\eta)\right) \} \ .
\label{bder}
\end{eqnarray}

The following discussion applies to both cases, {\it i.e.}, 
$\gamma^\mu$-coupling and derivative coupling.
The  matrix elements of the good component $j^+$ in Eq.(\ref{jvec})
are given by the sum of two terms:
$j^+=j^{+wf}+j^{+pair}$.
The first one, $j^{+wf}$, is the contribution of
the light-front wave-function, with $k^+$ integrated in the range
$0<k^+<P^+$ 
in Eq.(\ref{jvec}) \cite{pach97}. The second one  stems from the pair
term, $j^{+pair}$, obtained with Eqs.(\ref{pairv}, \ref{pairder}).
Each matrix element of the current acquires a
contribution from the pair term with the exception of  $j^+_{yy}$,
{\it i.e.}, $j^{+pair}_{yy} = 0$.  In other words, the longitudinal
zero mode contributes to  $j^+_{xx}$, $j^+_{zx}$ and $j^+_{zz}$.

The  angular condition  in the Cartesian spin basis is
\begin{eqnarray}
\Delta(q^2)=j^+_{yy}-j^+_{zz}=j^{+wf}_{yy}-j^{+wf}_{zz}-
j^{+pair}_{zz} \ = 0   \ .
\label{ca}
\end{eqnarray}
Since the matrix elements of the current $j^+_{ji}$ 
are obtained from a covariant model, \ $\Delta(q^2)$  
should indeed be identical zero. Below, cf.  Figs.4-5,  we will
explicitly confirm this in our calculations.
Considering only the residue at the pole of 
the spectator particle in the Cauchy integration over the longitudinal momentum
implies the violation of the angular condition.
 This has also been shown in a
recent numerical study of this model \cite{pach97}.

It is due to the omission of the pair
contribution to $j^+_{zz}$, {\it i.e.}, 
$j^{+pair}_{zz} = \ j^{+wf}_{yy}-j^{+wf}_{zz}$ \ . 
The  rotational symmetry of the "good" component of the 
current $j^+$ for a spin-one particle is restored, if the
pair creation process is carefully taken into account
in the computation of the Breit-frame matrix elements. 
The ``good'' component of the current receives a contribution 
from the pair term for $q^+=0$, and thus it is necessary to go beyond 
the naive light-front
integration in $k^-$. One should include the zero-mode
contribution to the current in order to calculate consistently
e.m. form-factors
in this reference frame.

\section{Numerical results and summary}

To illustrate the previous discussion, we 
compare  the numerical results for the matrix elements of $j^+$
as function of the momentum transfer $q^2$
obtained in the covariant formalism \cite{pach97} 
to the ones obtained in the light-front framework -- with
and without pair terms. 
We present the results
 for  $\gamma^\mu$-coupling and derivative coupling.
The parameters of the model  are taken from Ref.
\cite{pach97} and are  $m_v=$ 0.77 GeV, $m=$ 0.43 GeV and
$m_R=$ 1.8 GeV. 

Fig. 1 contains the results for $j^+_{xx}$.
For both couplings, sizeable deviations between covariant --
and light-front calculation, {\it not} taking into
account pair terms, show up.
Especially the derivative coupling yields a pronounced
difference. 
This fact is due to the higher powers of $k^-$ in the numerator of
Eq.(\ref{jvec}) as compared to the $\gamma^\mu$-coupling that has $k^-$ 
 up to the first order. 
For $q^2\rightarrow 0$, the  
zero mode contribution vanishes as $q^2$. 
We performed the
calculation of the pair current, $j^{+pair}_{xx}$, 
according to Eq.(\ref{pairv})  ($\gamma^\mu$-coupling)
and  Eq.(\ref{pairder}) (derivative coupling). In both
cases, including the pair amplitude in the light-front current 
reproduces the covariant results
within our numerical precision.

The same comparisons are made for the current component $j^+_{zx}$
in Fig.2 and for $j^+_{zz}$ in Fig.3. 
Here we also observe that the inclusion of pair production terms
in the light-front computation yields results in agreement with
the covariant ones. 
For small momentum transfer, the pair term in  $j^+_{zx}$
is proportional to $q_x$.
However, even the slope of the 
curve shows zero-mode effects in this limit. Increasing
the momentum transfer, the zero-mode contribution tends to be less important. 
In Figs.4-5, we show the angular condition,  
given by Eq.({\ref{ca}), for $\gamma^{\mu}$-coupling and derivative 
coupling, respectively. 
If the pair current is included in the light-front calculation,
this condition is fulfilled. This means that
$j^+_{yy}$ and $j^+_{zz}$ indeed coincide.
Omission of the light-front pair terms, however,
causes a  violation of the angular condition.
Note that even at $q_x=0$,
the zero-mode contribution is nonvanishing in $j^+_{zz}$. 
The numerical results not only determine the
magnitude of the effects arising from the pair current but also 
give further support to our analytical findings that the 
pair contribution is required to regain the covariant result.

In summary, using a fermion one-loop model
we have derived the contribution of the 
light-front pair amplitude 
to the e.m. current of a composite vector particle in
a reference frame where $q^+=0$. We have shown that the "good" 
component of the current $j^+$ gets contributions of the longitudinal
zero-momentum mode. Our discussion is based on a careful integration,
exploiting the  recently developed "dislocation of pole" technique,
in the  momentum-loop 
of the triangle diagram for the virtual photon absorption process. 
For two examples of 
the vector particle--fermion coupling, 
we have analytically and numerically calculated the pair current.

It  is shown that it is not possible to leave out that contribution
without violating Lorentz covariance and, consequently,
the angular condition in the Breit frame.

%%%%       \newpage 

\section*{Acknowledgments}
This work was supported in part by the Brazilian agencies 
CNPq and  FAPESP (contract 97/13902-8). 
It also was supported by 
 Deutscher Akademischer Austauschdienst  
and Funda\c c\~ao Coordena\c c\~ao de Aperfei\c coamento
de Pessoal de N\'\i vel Superior (Probral/CAPES/DAAD project 015/95).
T.F. thanks for  the hospitality of the Institute for Theoretical Physics
in Hannover. H.W.L.N. and P.U.S. thank for the hospitality of
Instituto Tecnol\'ogico de Aeron\'autica of S. Jos\'e dos Campos.

\begin{center}
{\bf APPENDIX}
\end{center}

In this Appendix, we  prove  Eq.(\ref{imm});
first we consider the case  $m=n=0$ and use induction in $N$.
Suppose that 
\begin{eqnarray}
I^{(N)}_{00}(\{ f_i\} )&=&
\frac12 \int 
\frac{d k^{+} d k^- }{\prod_{j=1}^N(P^+ -k^+)
(P^- - k^-- \frac{f_j -\imath \epsilon}{P^{+}-k^+})}
\nonumber \\
&=&
\imath \pi \sum_{j=1}^N\frac{ \ln(f_{j})}
{\prod_{k=1,k\neq j}^N(f_j - f_k)} \ ,
%\eqnum{(A.1)}
\label{a1}
\end{eqnarray}
is valid for some $N \ge 2$; below we will show its validity for $N+1$.

We begin by explicitly verifying the identity for  $N=2$
\begin{eqnarray}
&&I^{(2)}_{00}(\{ f_i\} )=
\nonumber \\
&=&\frac12 \int  
\frac{d k^{+} d k^- }{(P^++\delta -k^+)(P^+-k^+)
(P^- - k^-- \frac{f_1 -\imath \epsilon}{P^{+}+\delta-k^+})
(P^- - k^-- \frac{f_2 -\imath \epsilon}{P^{+}-k^+})} 
\nonumber \\
&=& \imath \pi \frac{\ln(f_1)-\ln(f_2)}{f_1 - f_2} \ .
%\eqnum{(A.2)}
\label{a2}
\end{eqnarray}
Next we use that:
\begin{eqnarray}
&&\frac{1 }{(P^+ -k^+)
(P^- - k^-- \frac{f_N -\imath \epsilon}{P^{+}-k^+})}
\times
\frac{1}{(P^+-k^+)
(P^- - k^-- \frac{f_{N+1} -\imath \epsilon}{P^{+}-k^+})} 
= \frac{1}{f_{N}-f_{N+1}}
\nonumber \\
&&\times\left(\frac{1 }{(P^+ -k^+)
(P^- - k^-- \frac{f_N -\imath \epsilon}{P^{+}-k^+})}- 
\frac{1}{(P^+-k^+)
(P^- - k^-- \frac{f_{N+1} -\imath \epsilon}{P^{+}-k^+})} 
\right) .
%\eqnum{A.3}
\label{a3}
\end{eqnarray}
With the aid of Eq.(\ref{a2}), we write:
\begin{eqnarray}
& &I^{(N+1)}_{00}(\{ f_i\} )=
\nonumber \\
&=&\frac{i\pi}{f_{N}-f_{N+1}}\left[
 \sum_{j=1}^N\frac{ \ln(f_{j})}
{\prod_{k=1,k\neq j}^N(f_j - f_k)} -
\sum_{j=1,j\neq N}^{N+1}\frac{ \ln(f_{j})}
{\prod_{k=1,k\neq j, k\neq N}^N(f_j - f_k)}
\right] 
\nonumber \\
&=&\frac{i\pi}{f_{N}-f_{N+1}}\left[
 \sum_{j=1,j\neq N+1}^{N+1}\frac{ \ln(f_{j})}
{\prod_{k=1,k\neq j, k\neq N+1}^{N+1}(f_j - f_k)} -
\sum_{j=1,j\neq N}^{N+1}\frac{ \ln(f_{j})}
{\prod_{k=1,k\neq j, k\neq N}^N(f_j - f_k)}
\right] 
\nonumber \\
&=&\frac{i\pi}{f_{N}-f_{N+1}}\left[
 \sum_{j=1,j\neq N+1}^{N+1}\frac{ \ln(f_{j})(f_j-f_{N+1})}
{\prod_{k=1,k\neq j}^{N+1}(f_j - f_k)} -
\sum_{j=1,j\neq N}^{N+1}\frac{ \ln(f_{j})(f_j-f_{N})}
{\prod_{k=1,k\neq j}^N(f_j - f_k)}
\right] 
\nonumber \\
&=&\frac{i\pi}{f_{N}-f_{N+1}}
 \sum_{j=1,}^{N+1}\frac{ \ln(f_{j})(f_j-f_{N+1}-f_j+f_N)}
{\prod_{k=1,k\neq j}^{N+1}(f_j - f_k)}  
\nonumber \\
&=&i\pi
 \sum_{j=1}^{N+1}\frac{ \ln(f_{j})}
{\prod_{k=1,k\neq j}^{N+1}(f_j - f_k)} 
\ . 
%\eqnum{A.4}
\label{a4}
\end{eqnarray}

Secondly, we prove Eq.(\ref{imm}) by induction in $m$. 
Suppose that 
\begin{eqnarray}
I^{(N)}_{mm}(\{ f_i\} )&=&\int
\frac{d k^{+} d k^- }{2} 
\frac{(k^-)^m(P^+-k^+)^m}{\prod_{j=1}^N(P^+ -k^+)
(P^- - k^-- \frac{f_j -\imath \epsilon}{P^{+}-k^+})} 
\nonumber \\
&=&\imath \pi \sum_{j=1}^N\frac{ (-f_{j})^m\ln(f_{j})}
{\prod_{k=1,k\neq j}^N(f_j - f_k)} \ , 
%\eqnum{A.5}
\label{a5}
\end{eqnarray}
is valid for $m$, then consider
\begin{eqnarray}
&&I^{(N)}_{m+1\ m+1}(\{ f_i\} )=
\int
\frac{d k^{+} d k^- }{2}\times
\nonumber \\ 
&&\frac{(k^--P^-)(k^-)^{m}(P^+-k^+)^{m+1}+P^-(k^-)^{m}(P^+-k^+)^{m+1}}
{\prod_{j=1}^N(P^+ -k^+)
(P^- - k^-- \frac{f_j -\imath \epsilon}{P^{+}-k^+})} 
 \ . 
%\eqnum{A.6}
\label{a6}
\end{eqnarray}
The second term in the integrand of Eq.(\ref{a6}) vanishes due to
 $I^{(N)}_{mn}({f_i})=0$ for $m \ <\ n$.  Then we obtain
\begin{eqnarray}
& &I^{(N)}_{m+1\ m+1}(\{ f_i\} )=
\nonumber \\
&=&-\int
\frac{d k^{+} d k^- }{2} 
\frac{\left( (P^--k^-)(P^+-k^+)-f_1+f_1 \right)(k^-)^{m}(P^+-k^+)^{m}}
{\prod_{j=1}^N(P^+ -k^+)
(P^- - k^-- \frac{f_j -\imath \epsilon}{P^{+}-k^+})} 
\nonumber \\ 
&=&-\int
\frac{d k^{+} d k^- }{2} 
\frac{(k^-)^{m}(P^+-k^+)^{m}}
{\prod_{j=1, j\neq 1}^N(P^+ -k^+)
(P^- - k^-- \frac{f_j -\imath \epsilon}{P^{+}-k^+})}
\nonumber \\
&&-f_1\int
\frac{d k^{+} d k^- }{2} 
\frac{(k^-)^{m}(P^+-k^+)^{m}}
{\prod_{j=1}^N(P^+ -k^+)
(P^- - k^-- \frac{f_j -\imath \epsilon}{P^{+}-k^+})}
\nonumber \\ 
&=&-\imath \pi \left[ \sum_{j=1, j\neq 1}^N\frac{ (-f_{j})^m\ln(f_{j})}
{\prod_{k=1,k\neq j,k\neq 1}^N(f_j - f_k)} +
\sum_{j=1}^N\frac{ f_1(-f_{j})^m\ln(f_{j})}
{\prod_{k=1,k\neq j}^N(f_j - f_k)} 
\right]
\nonumber \\ 
&=&-\imath \pi \left[ \sum_{j=1}^N\frac{(f_j-f_1) (-f_{j})^m\ln(f_{j})}
{\prod_{k=1,k\neq j}^N(f_j - f_k)} +
\sum_{j=1}^N\frac{ f_1(-f_{j})^m\ln(f_{j})}
{\prod_{k=1,k\neq j}^N(f_j - f_k)} 
\right]
\nonumber \\ 
&=&\imath \pi  \sum_{j=1}^N\frac{(-f_{j})^{m+1}\ln(f_{j})}
{\prod_{k=1,k\neq j}^N(f_j - f_k)}  
, 
%\eqnum{A.7}
\label{a7}
\end{eqnarray}
which completes the proof.

%% \begin{references}

%%%% FIGURES / EPS %%%%%%

\begin{figure}[h]
\vspace{15.0cm}
\caption{Current component  $j^+_{xx}$ as a function of $|q^2|$.
Results for  $\gamma^\mu$-coupling: 
covariant and  light-front including pair current (full circles),
light-front without pair current (dotted curve).
Results for derivative coupling:
covariant and light-front including pair current
(solid curve), light-front without pair current 
(dashed curve). The results for the covariant calculation and 
for the light-front calculation including pair current contribution coincide.}
\label{fig1} 
\includegraphics{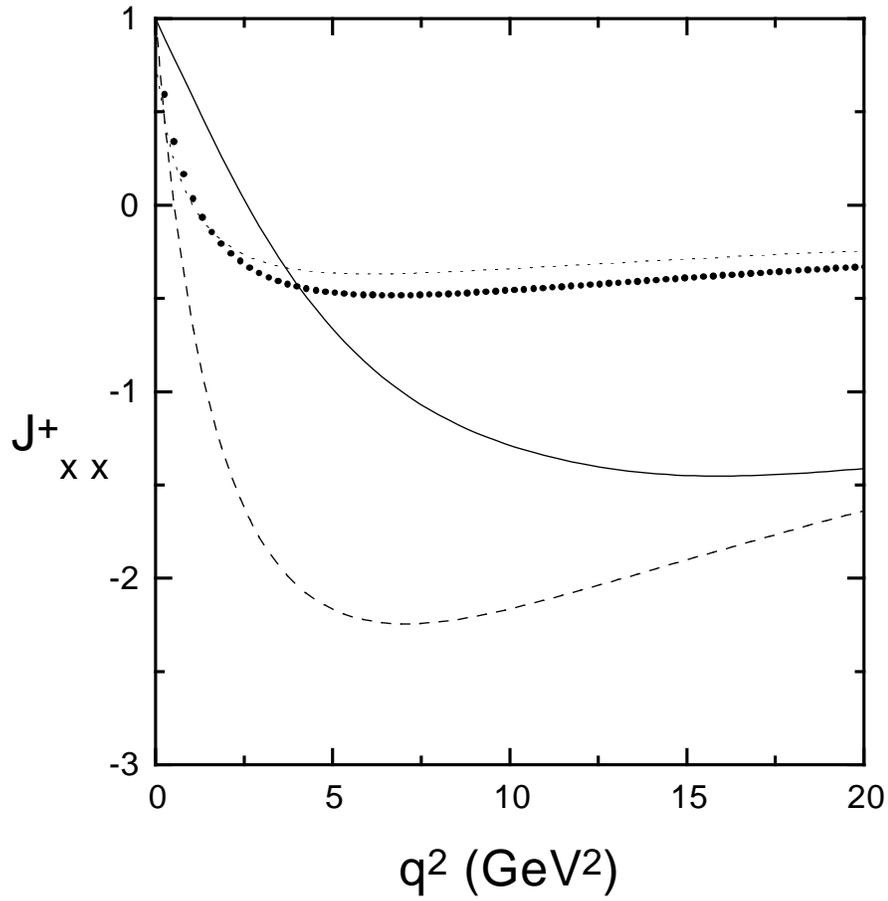} 
\end{figure}

\newpage 
.
\vskip 2cm

\begin{figure}[h]
\begin{center}
\vspace{15.0cm}
\caption{Current component $j^+_{zx}$ as a function of $|q^2|$.
Curves labelled as in Fig. 1.}
\label{fig2} 
\includegraphics{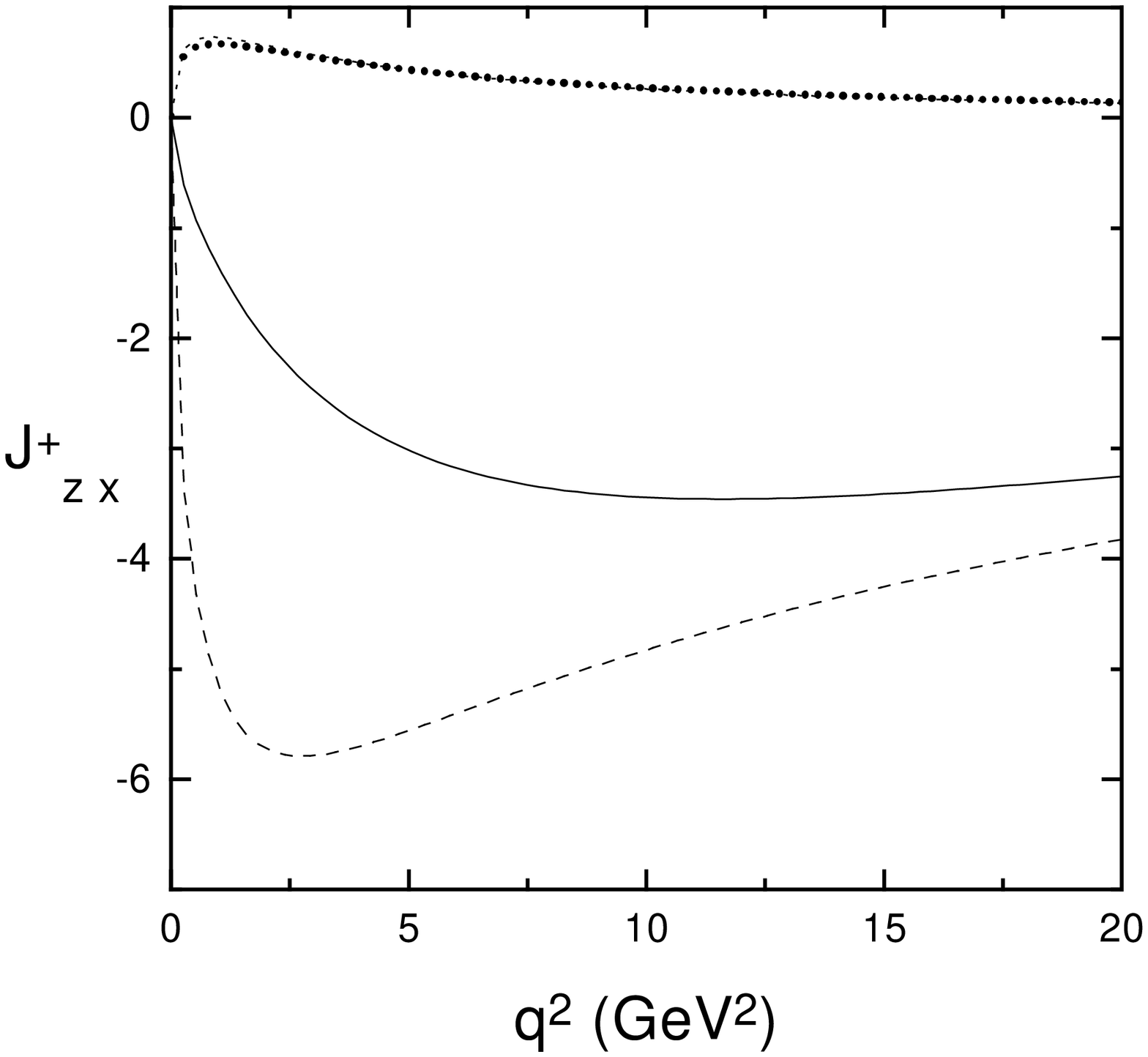} 
\end{center}
\end{figure}

\newpage 
.
\vskip 2cm
\begin{figure}[h]
\vspace{15.0cm}
\caption{Current component $j^+_{zz}$ as a function of $|q^2|$.
Curves labelled as in Fig. 1.}
\label{fig3} 
\includegraphics{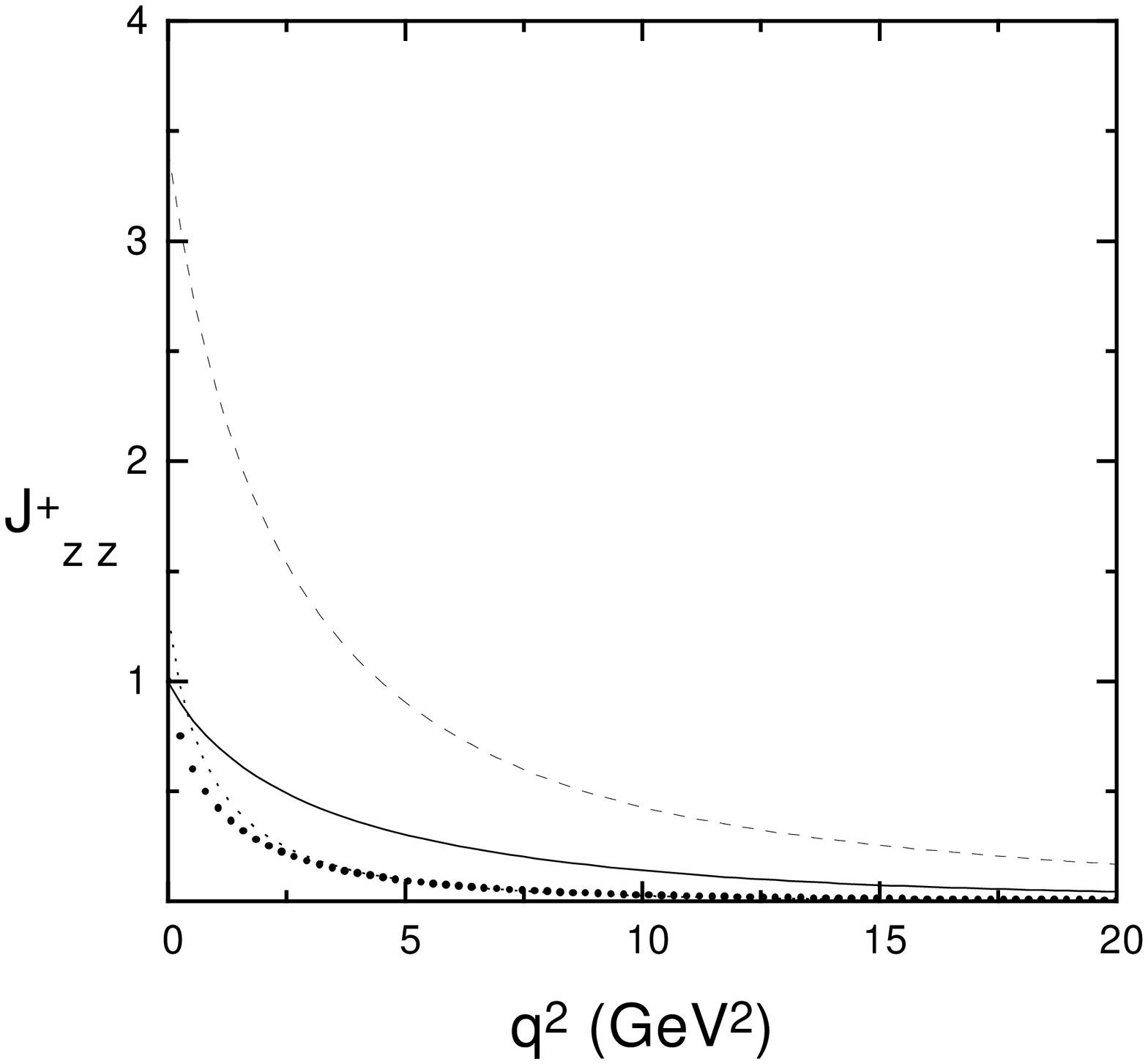} 
\end{figure}

\newpage 
.
\vskip 2cm
\begin{figure}[h]
\vspace{15.0cm}
\caption{$\Delta$ as a function of $|q^2|$
for  $\gamma^{\mu}$-coupling. 
Dashed:  light-front without pair terms, dotted:  pair terms,
full circles: light-front including pair terms, solid: covariant 
calculation.}
\label{fig4} 
\includegraphics{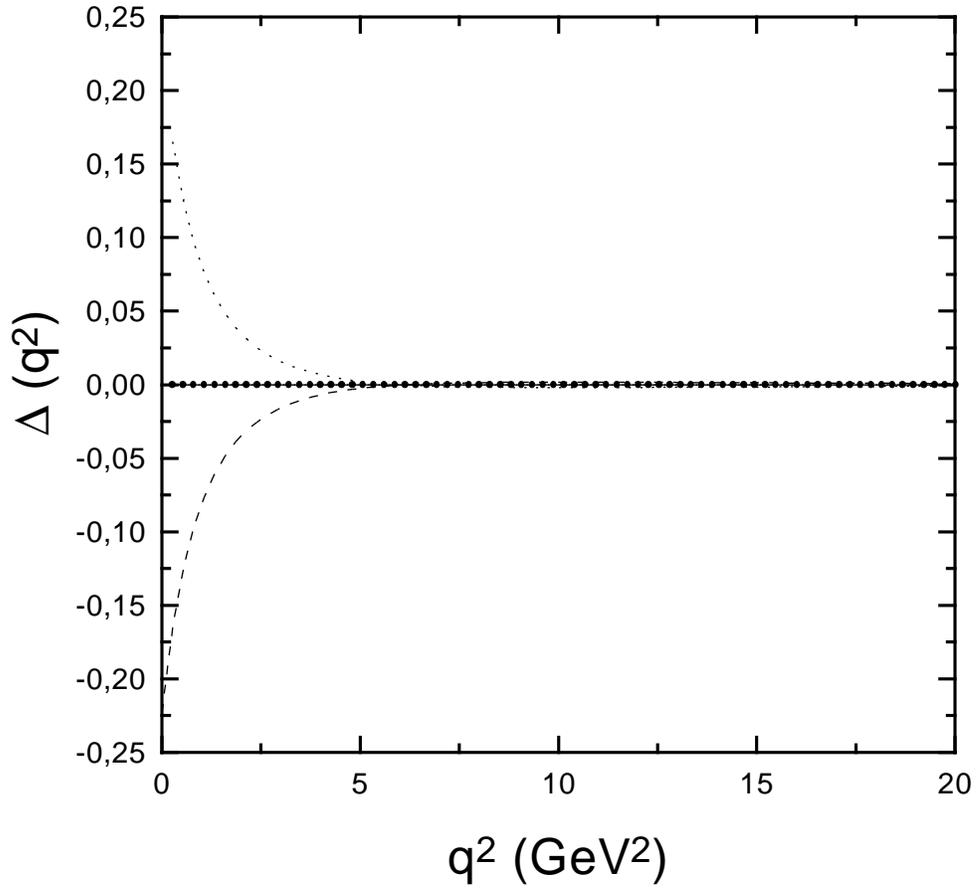} 
\end{figure}

\newpage 
.
\vskip 2cm
\begin{figure}[h]
\vspace{15.0cm}
\caption{$\Delta$ as a function of  $|q^2|$
for  derivative coupling; labelling as in Fig.4.} 
\label{fig5} 
\includegraphics{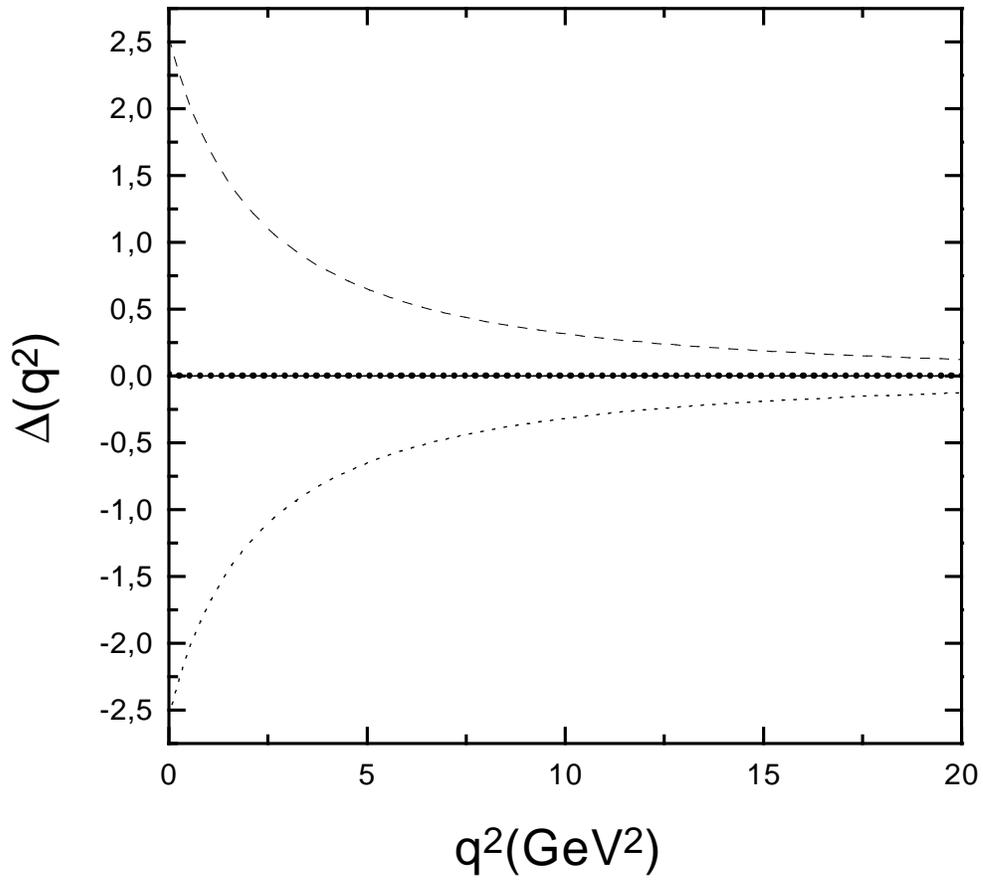} 
\end{figure}

\end{document}